\documentclass[aps,prl,twocolumn,floatfix,showpacs]{revtex4}
\pdfoutput=1
\usepackage[pdftex]{color}
\usepackage{graphicx}
\usepackage{amsmath,amssymb,bm}
\usepackage{color}

\def\rnum#1{\expandafter{%
\romannumeral #1}}
\def\Rnum#1{\uppercase\expandafter{%
\romannumeral #1}}

\begin{document}

\title{
Chiral order and electromagnetic dynamics in one-dimensional multiferroic cuprates
}
\author{Shunsuke Furukawa}
\altaffiliation{Present address: Department of Physics, University of Toronto, Toronto, Ontario M5S 1A7, Canada}
\affiliation{Condensed Matter Theory Laboratory, RIKEN, Wako, Saitama 351-0198, Japan}
\author{Masahiro Sato}
\affiliation{Condensed Matter Theory Laboratory, RIKEN, Wako, Saitama 351-0198, Japan}
\author{Shigeki Onoda}
\email{s.onoda@riken.jp}
\affiliation{Condensed Matter Theory Laboratory, RIKEN, Wako, Saitama 351-0198, Japan}
\date{\today}


\begin{abstract}
We show by unbiased numerical calculations that the ferromagnetic nearest-neighbor exchange interaction stabilizes a vector spin chiral order against the quantum fluctuation in a frustrated spin-$\frac{1}{2}$ chain relevant to multiferroic cuprates, LiCu$_2$O$_2$ and LiCuVO$_4$. 
Our realistic semi-classical analyses for LiCu$_2$O$_2$  resolve controversies on the helical magnetic structure and unveil the pseudo-Nambu-Goldstone modes as the origin of experimentally observed electromagnons. 
\end{abstract}
\maketitle

\begin{figure}[b]
\begin{center}
\includegraphics[width=0.9\columnwidth]{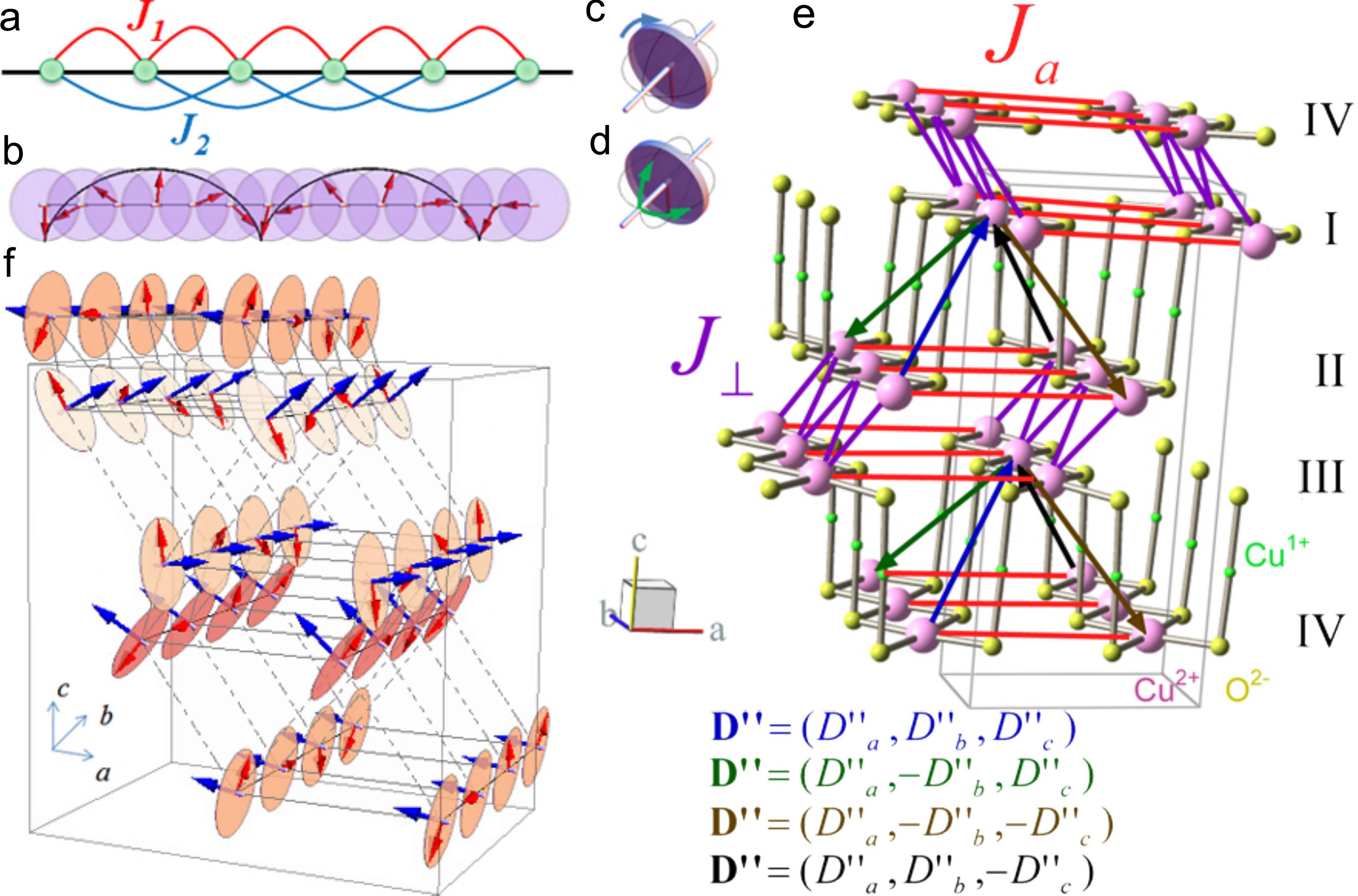}
\end{center}
\caption{
(Color)
(a) A 1D frustrated spin model with $J_1$ and $J_2 (>0) $ being the nearest-neighbor and second-neighbor exchange couplings.
(b) A spin cycloid in the propagation direction of the $J_1$-$J_2$ spin chain. Spins rotate within the common plane represented by the disks.
(c,d) Three Nambu-Goldstone modes in the SU(2)-symmetric model; a phason [(c)] and two rotation modes of the spiral plane [(d)].
(e) Cu$^{2+}$, Cu$2^{1+}$ and O$^{2-}$ ions in LiCu$_2$O$_2$, and inter-chain Heisenberg exchange ($J_{a,\perp}$) and inter-layer Dzyaloshinskii-Moriya interactions. On the arrows with four different colors, four distinct DM vectors ($\bm{D}"$) are assigned.
(f) Hypothetical magnetic structure (red arrows) and lowest-energy spin-excitation mode (rotations around blue arrows) in the THz range for LiCu$_2$O$_2$. 
}
\label{fig:cycloid}
\end{figure}

The parity symmetry may be broken spontaneously in magnets~\cite{yoshimori,nagamiya,villain:78} even if the original crystal structure preserves it.
It occurs, for instance, when the vector chirality \cite{villain:78} 
$\bm{\kappa}_{\bm{r},\bm{r}'}=\langle \bm{S}_{\bm{r}}\times\bm{S}_{\bm{r}'}\rangle$ of neighbouring spins $\bm{S}_{\bm{r}}$ and $\bm{S}_{\bm{r}'}$ acquires a nonzero macroscopic average in geometrically frustrated magnets [Fig.~\ref{fig:cycloid}(a)].
The helical magnetism~\cite{yoshimori,nagamiya} in which the spins align in a helix with a particular handedness is a typical example. 
This issue of the spin chiral order~\cite{villain:78} and the associated electromagnetic excitations have been highlighted by a recently discovered multiferroic behavior, a ferroelectricity induced by a spin cycloid \cite{kimura:03,cheong}. 
In spite of theoretical advances in the understanding of the static magnetoelectric effect \cite{knb,sergienko:06,vanderbilt}, 
the dynamical effects \cite{pimenov:06,kbn,aguilar:09,huvonen:09} remain controversial.

In helical magnets, three Nambu-Goldstone modes appear in principle~\cite{nagamiya,yoshimori} since the SU(2) symmetry of the spins are fully broken. 
In the case of the spin cycloid [Fig.~\ref{fig:cycloid}(b)], one corresponds to a phason describing an infinitesimal change of the pitch of the cycloid 
[Fig.~\ref{fig:cycloid}(c)]. 
The other two represent infinitesimal rotations of the cycloid plane [Fig.~\ref{fig:cycloid}(d)].  
Because of magnetic anisotropy, they usually acquire an energy gap and can then be probed with the antiferromagnetic resonance \cite{vorotynov:98}. 
These modes may also be excited by the electric component of light through the magnetoelectric coupling. 
Recent THz spectroscopy experiments on $R$MnO$_3$ \cite{pimenov:06,aguilar:09} have demonstrated that this electromagnon spectrum grows below the ferroelectric transition temperature. 
However, because of the large GdFeO$_3$-type distortion, the observed spectrum is dominated by high-energy magnons at the Brillouin-zone boundary \cite{aguilar:09} through the magnetostriction mechanism \cite{jonh}. 
The roles of the Nambu-Goldstone modes remain open \cite{kbn}.

\begin{figure*}
\begin{center}
\includegraphics[width=0.9\textwidth]{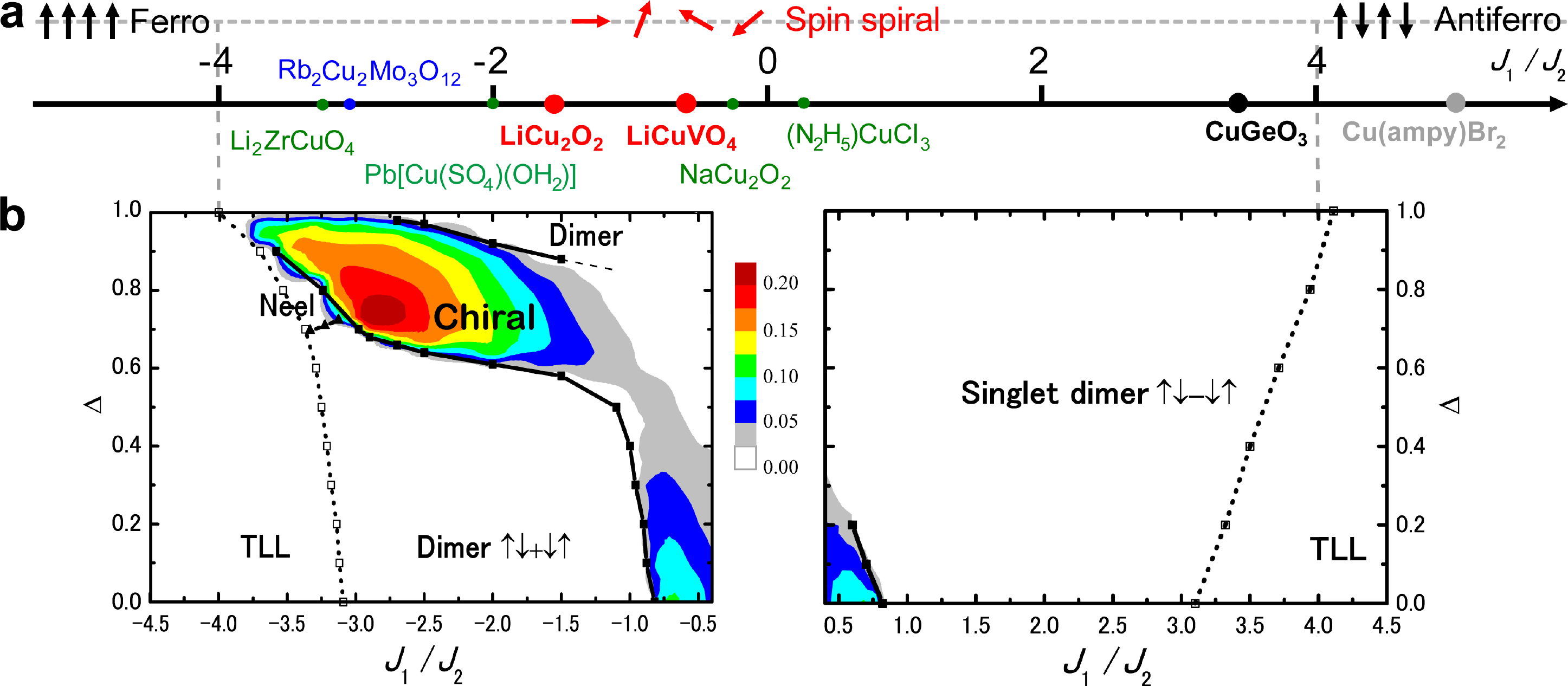}
\end{center}
\caption{
(Color) Ground-state phase diagram of the spin chain model $H_{1D}$. 
(a)
The classical phase diagram and relevant Q1D spin-$\frac12$ materials. 
Estimates of $J_1/J_2$ for materials are taken from Refs.~\onlinecite{masuda:05,drechsler:07,hase:04}. 
The materials shown in green exhibit an antiferromagnetic LRO, but their detailed magnetic structures have not been established yet. 
In Rb$_2$Cu$_2$Mo$_3$O$_{12}$, a magnetic LRO has not been detected down to 2 K~\cite{hase:04}. Usually, $1-\Delta\lesssim0.05$ in cuprates.
(b)
The quantum phase diagram and the map of the chiral order parameter $\kappa^z$ for $S=\frac12$. The horizontal axis $J_1/J_2$ is the same as in the panel (a).
The boundary between the dimer phase and the chiral phase in the case of $J_1>0$ agrees with the previous numerical study~\cite{hikihara:01}. 
For $|J_1|/J_2\gtrsim4$, Tomonaga-Luttinger liquid (TLL) phases appear.
Detailed methods of identifying each phase and the phase boundaries are explained in Supplementary Material. 
It was difficult to determine the phase boundary for small $|J_1|/J_2$ (dashed line).
}
\label{fig:chirality}
\end{figure*}

The ferroelectricity associated with the spin cycloid has also been found in quasi-one-dimensional (Q1D) cuprates, LiCu$_2$O$_2$ \cite{park:07} and LiCuVO$_4$ \cite{naito:07}. 
These spin-chain compounds are advantageous in observing the lowest-energy magnons in the optical spectrum since the crystal structure is less distorted along the chains. 
In particular, in LiCu$_2$O$_2$, the contribution from the zone-boundary magnons along the chain is prohibited by the symmetry. 
However, the understanding of experimental findings on the Q1D multiferroic cuprates remains controversial, 
including the magnetic ordering pattern \cite{masuda:05,gippius:04,seki:08,huvonen:09,kobayashi:09} and the dynamical properties \cite{huvonen:09} in LiCu$_2$O$_2$. 
Furthermore, it is by far nontrivial from the theoretical viewpoint whether the chiral LRO parasitic to the helical magnetism can be stabilized against quantum fluctuations. 
In fact, strong quantum fluctuations in one dimension often lead to a valence-bond order accompanied by a moderately large gap in the spin excitations~\cite{white:96,hikihara:01}. 
This spin gap can prevail over weak inter-chain couplings and prevent the helical magnetic long-range order (LRO), as in CuGeO$_3$ \cite{hase:93}. 
A possibility that the chiral ordered phase can appear for a weak easy-plane anisotropy in the frustrated spin-$\frac12$ chain has been addressed \cite{nersesyan:98,furukawa:08}, and conclusive calculations are called for. 

In this Letter, we develop a comprehensive theory for Q1D multiferroic cuprates. 
We uncover that the ferromagnetic nearest-neighbor exchange coupling $J_1$ stabilizes a chiral order in the frustrated spin-$\frac12$ chain under weak easy-plane anisotropy, in sharp contrast to the case of the antiferromagnetic $J_1$ [Fig.~2(b)]. 
With weak three-dimensional (3D) couplings, this roughly controls whether the ground state exhibits the helical magnetic order or the Neel/dimer order. 
Our phase diagram is useful for classifying several Q1D cuprates \cite{hase:04,drechsler:07} [Fig.~2]. 
We also theoretically clarify the magnetic ordering structure and the electromagnetic excitation spectra in LiCu$_2$O$_2$, 
which show overall agreements with experiments \cite{masuda:05,gippius:04,park:07,seki:08,kobayashi:09,huvonen:09}. 
These agreements give evidence that weak inter-chain couplings and magnetic anisotropy allow the electric component of light to excite otherwise zero-energy Nambu-Goldstone modes.

We first reveal the origin of the vector chiral order underlying the spin spiral in LiCuVO$_4$ and LiCu$_2$O$_2$. 
At higher temperatures than the inter-chain couplings of order of the N\'eel temperature (2.5~K and 24~K for LiCuVO$_4$ and LiCu$_2$O$_2$, respectively), they are described in terms of a one-dimensional (1D) spin-$\frac12$ model [Fig.~1(a)], 
\begin{equation}
H_{1D}=\sum_{n=1,2}J_n\sum_j\left(S^x_jS^x_{j+n}+S^y_jS^y_{j+n}+\Delta S^z_jS^z_{j+n}\right),
\end{equation}
($J_2>0$) with the electronic spin $\bm{S}_j=(S^x_j,S^y_j,S^z_j)$ at the Cu$^{2+}$ site $j$ in the chain, and the small symmetric easy-plane exchange anisotropy $0<1-\Delta\ll1$. In the classical limit, a helical magnetic ground state is realized for $|J_1|/J_2<4$ [Fig.~2(a)], with a finite uniform vector spin chirality $\langle\kappa^z_{j,j+1}\rangle=\kappa^z$.
In the quantum spin-$\frac12$ case, this chiral order tends to yield to valence bond orders~\cite{white:96,hikihara:01}.
However, our previous finite-size calculation of the spin Drude weight, supplemented by the bosonization analysis, has suggested that the chiral phase might survive in a wider range for ferromagnetic $J_1$ \cite{furukawa:08}.  
To precisely determine the ground-state phase diagram in a conclusive manner, we have performed numerical calculations based on the time evolving block decimation algorithm for an infinite system (iTEBD)~\cite{vidal:07}. 
Figure~2(b) presents the global phase diagram and a profile of the chiral order parameter $\kappa^z$ in the space of $J_1/J_2$ and $\Delta$. 
It confirms that the chiral phase extends over a wide region for ferromagnetic $J_1<0$. In particular, the chiral order is pronounced by small magnetic anisotropy $1-\Delta>0$ and a moderate value of $|J_1|/J_2$ that roughly correspond to the multiferroic cuprates. This chiral LRO is stable up to a close vicinity of the $SU(2)$-symmetric case $\Delta=1$, where it is replaced by a dimer order. It is also replaced with another dimer order with a unit of $|\!\!\uparrow\downarrow\rangle+|\!\!\downarrow\uparrow\rangle$ by stronger anisotropy $0\le \Delta \lesssim 0.6$. On the other hand, for antiferromagnetic $J_1>0$, the singlet-dimer order accompanied by the spin gap is robust   \cite{white:96,hikihara:01}.

The above results clearly explain why the Q1D helical magnets or the candidate materials are found dominantly on the ferromagnetic side of $J_1$ while the materials located on the antiferromagnetic side of $J_1$ usually show a spin-singlet dimer order or a collinear antiferromagnetic LRO. 
In real materials having weak inter-chain couplings, the helical spin quasi-LRO in the chiral phase \cite{nersesyan:98} readily evolves into a genuine helical LRO. 
Even when the magnetic anisotropy is so small that a material with $J_1<0$ is located in a narrow dimer phase near $\Delta=1$, the spin gap is orders of magnitude smaller than for $J_1>0$. 
Therefore, the weak 3D couplings may close the spin gap and recover the helical magnetic LRO. 
Once the helical magnetic LRO readily occurs at a finite temperature on the ferromagnetic side of $J_1$, as in the case of LiCu$_2$O$_2$,
then the static and dynamical properties can be calculated through semi-classical analyses \cite{nagamiya}. In the rest of this Letter, constructing a realistic spin model, we will give a theoretical explanation of overall experimental findings on LiCu$_2$O$_2$.

The material contains four $J_1$-$J_2$ spin chains [Fig.~1(a)], each of which lies in one of four stacked $ab$ layers (I to IV) in the unit cell [Fig.~1(e)]. 
Neutron-scattering experiments \cite{masuda:05} have unveiled a magnetic Bragg peak at an incommensurate wavevector $(0.5,0.173,0)$ in the reciprocal lattice unit, and the Q1D spin-wave dispersion along the $b$ axis. 
The observation of the ferroelectric polarization $\bm{P}\parallel c$ \cite{park:07,seki:08} and the polarized neutron-scattering experiment~\cite{seki:08} indicate the emergence of the uniform $bc$ spin-spiral component in the lowest-temperature phase. Then, a minimal spin model should include the interactions shown in Fig.~1(e). 
In each layer, the adjacent spin chains along the $a$ axis are coupled via the Heisenberg exchange interaction with moderate coupling strength $J_a$ [Fig.~1(e)]. Two pairs of the adjacent layers, II and III, and IV and I, are also coupled via the nearest-neighbor inter-chain Heisenberg exchange coupling $J_\perp$ in a zigzag manner [Fig.~1(e)]. On the other hand, a small Heisenberg exchange coupling $J'$ (not shown) between the other two pairs of layers, I and II, and III and IV, induces an incommensurability along the $a$ axis, i.e., $Q_a\ne\pi/a$, which disagrees with  the experiment~\cite{masuda:05,kobayashi:09,seki:08}. To stabilize the experimentally reported uniform $bc$ spiral component \cite{park:07,seki:08,huvonen:09,kobayashi:09} and to pin the commensurability along the $a$ axis, the $a$ component of the Dzyaloshinskii-Moriya (DM) vectors for the inter-layer DM interactions between I and II and between III and IV is indispensable. This can be provided through the second-neighbor bonds characterized by the four DM vectors $\bm{D}''$, which are specified by the amplitudes of the $a$, $b$, and $c$ components $D_a''$, $D_b''$, and $D_c''$ [Fig.~1(e)] from the symmetry argument. For simplicity, we take $J'=D''_b=D''_c=0$. The intra-chain DM vectors specified by the amplitude $D$ and the angle $\theta$ are uniform within each layer, but the symmetries of the crystal require that their directions alternate in the layer index; $\bm{D}_\mathrm{I}=D(\cos\theta,0,\sin\theta)$, $\bm{D}_\mathrm{II}=D(-\cos\theta,0,\sin\theta)$, $\bm{D}_\mathrm{III}=D(\cos\theta,0,-\sin\theta)$, and $\bm{D}_\mathrm{IV}=D(-\cos\theta,0,-\sin\theta)$. The classical analysis of this model Hamiltonian shows that the intra-chain DM interaction rotates the cycloid planes about the $b$ axis from the $bc$ plane by different angles $\varphi_1$, $\varphi_2$, $-\varphi_1$, and $-\varphi_2$ for the layers I, II, III, and IV, respectively. With the Bragg-peak position $(0.5,0.173,0)$ \cite{masuda:05}, a choice of $(\varphi_1,\varphi_2)/(2\pi)=(-0.08,-0.02)$ reproduce the $^7$Li nuclear magnetic resonance spectra \cite{gippius:04,kobayashi:09} (Supplementary Figure~S4). Then, the exchange and DM coupling parameters are determined to reproduce the ordering wavevector $\bm{Q}$, the rotation angles $(\varphi_1, \varphi_2)$ of the spiral planes, the anisotropy in the antiferromagnetic resonance spectra \cite{huvonen:09}; $(J_1,J_2,J_a,J_\perp,D,D''_a)=(-11.3,5.9,1.1,0.08,0.37,0.26)$~meV and $\theta/(2\pi)=0.43$. It is noteworthy that our choice $(J_1, J_2,J_a,J_\perp) $ is within a reasonable range with respect to previous analyses of inelastic neutron-scattering experiments \cite{masuda:05}. The obtained magnetic structure is presented in Fig.~1(f).

\begin{figure}[t]
\begin{center}
\includegraphics[width=0.9\columnwidth]{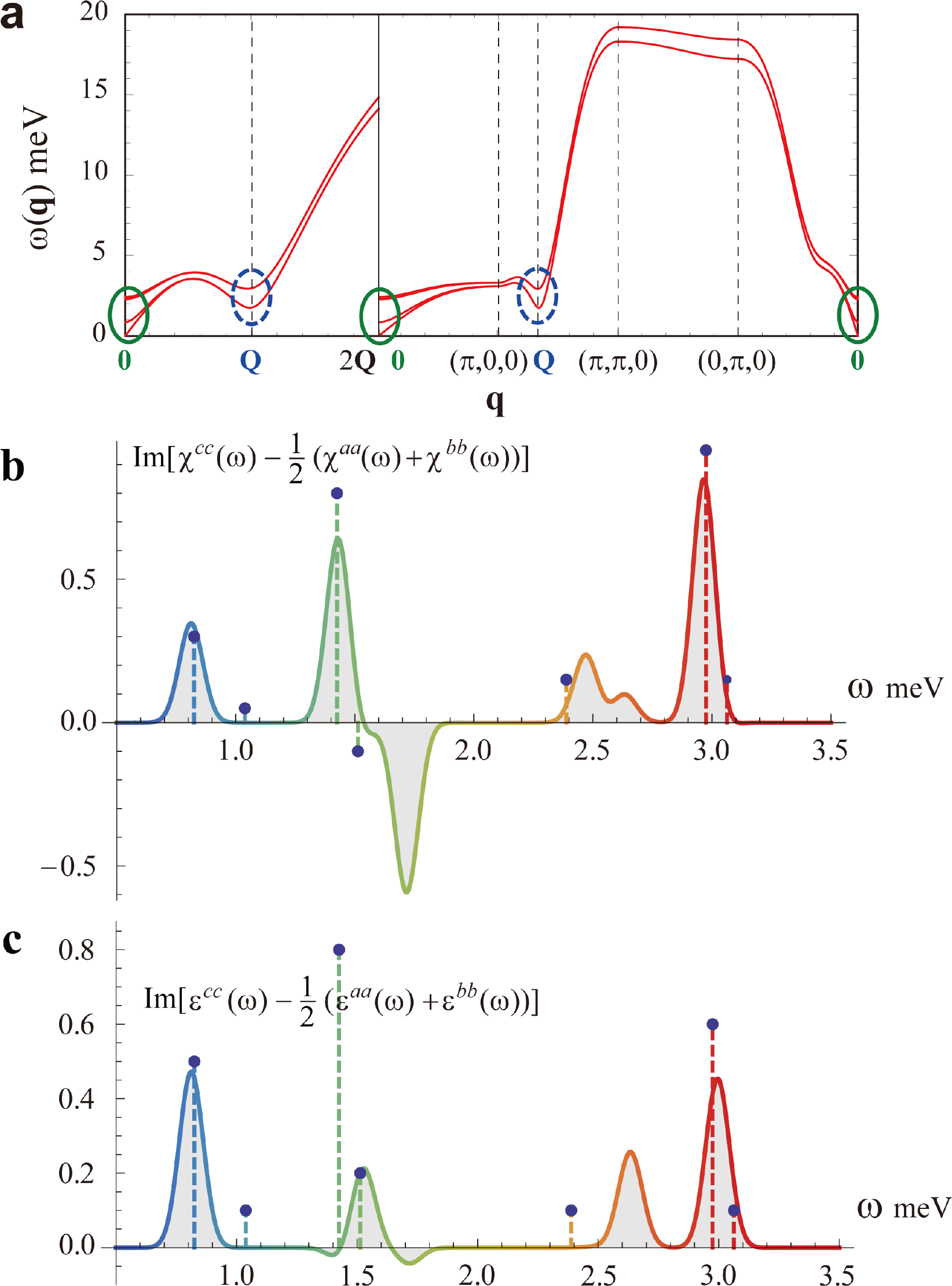}
\end{center}
\caption{
(Color online)
Semi-classical analyses for LiCu$_2$O$_2$. 
(a) Spin-wave dispersions without taking account of the bilinear coupling among $\bm{q}=0,\pm\bm{Q}$ modes. This coupling only modifies the modes around $\bm{q}=0$ and $\pm\bm{Q}$, marked with green and blue (dashed) circles.
The dispersion along the $c$ axis is quite small, since $J_\perp, D''_a< J_a \ll |J_1|, J_2$.
(b,c) Anisotropy in the electromagnetic absorption spectra due to the magnetic and electric components of light, 
in units of $(g\mu_B)^2$ meV$^{-1}$ and cm$^{-1}$, respectively. 
The spectrum consists of eleven gapped modes after the reconstruction of the $\bm{q}=0,\pm\bm{Q}$ modes due to their bilinear coupling; 
$\omega=0.812$, 1.431, 1.525, 1.717 (doubly degenerate), 2.469, 2.634, 2.977, 2.983, 2.991 (doubly degenerate) meV. 
Inclusion of even higher harmonics such as $\bm{q}=\pm2\bm{Q}$ modes increases the number of the modes and eventually broadens the peaks. 
The series of $\delta$-function peaks are replaced with the sum of Gaussians broadened with the width 0.05 meV.
Points with dashed lines denote the integrated intensity experimentally observed in the THz spectroscopy \cite{huvonen:09} in unit of cm$^{-1}$.
We note that the tilting of the spiral planes crucially changes the signals of electromagnons. 
}
\label{fig:magnon}
\end{figure}

The associated Q1D spin-wave dispersions are shown in Fig.~3(a). Magnetic anisotropy has produced the energy gap in otherwise zero-energy Nambu-Goldstone modes at $\bm{q}=\pm\bm{Q}$. 
Since four spins exist in the unit cell, there appear an acoustic branch and three optical branches. The splitting of the four branches is pronounced at the dispersion minima, $\bm{q}=0$ and $\bm{q}=\pm\bm{Q}$, where we further take account of the bilinear mixing of $\bm{q}=0$ and $\bm{q}=\pm\bm{Q}$ magnons through the degenerate perturbation theory of inter-layer and DM couplings. 
Figure~3(b) shows the anisotropy of the obtained antiferromagnetic resonance spectra $\mathrm{Im}[\chi^{cc}(\omega)-(\chi^{aa}(\omega)+\chi^{bb}(\omega))/2]$, which remarkably agrees with the experimental result \cite{huvonen:09}.  The lowest-energy mode at $\omega\sim 0.8$ meV is associated with a linear combination of the intra-layer phason modes where the change of the pitch of the cycloid in the layers I and II is opposite to that in the layers III and IV [Fig.~1(f)]. It is also accompanied by a small fraction of $2\bm{Q}$ oscillation. The acoustic mode was observed at $\sim30$~GHz~$\sim1$~cm$^{-1}\sim0.12$~meV with the electron spin resonance experiment \cite{vorotynov:98}, which is much smaller than the THz frequency of our interest.

Our analyses of magnetoelectric couplings (based on Ref.~\onlinecite{jonh}) show that the dominant contribution to the dielectric absorption spectra arises from the vector chirality on the bonds connecting the adjacent layers (Supplementary Material). 
The dominance of the chirality contribution over the magnetostriction sharply contrasts with the case of $R$MnO$_3$ \cite{aguilar:09}, and is ascribed to the fact that the unit cell contains only a single spin in the direction of the ordering wavevector.
Through this magnetoelectric coupling, most of the spin-wave modes are electric-dipole active.  
Since main peaks observed in the THz spectroscopy \cite{huvonen:09} are much sharper than the magnon bandwidth of order of $|J_1|$ and $J_2$, we take account only of the one-magnon contributions.  
The anisotropy of thus determined dielectric functions, $\mathrm{Im}[\varepsilon^{cc}(\omega)-(\varepsilon^{aa}(\omega)+\varepsilon^{bb}(\omega))/2]$, shows a reasonable agreement with the experiment \cite{huvonen:09} [Fig.~3(c)]. 
As expected, the lowest-energy mode described in Fig.~1(f) has significant amplitude as well.
We stress that excitations induced by electric and magnetic components of light appear with comparable amplitude in this system. 
This is unique to spin-$\frac12$ systems, where the magnetostriction contribution to the optics is absent and the magnetoelectric coupling is controlled by quite a small ratio of the relativistic spin-orbit coupling to the on-site Coulomb repulsion.

Dynamical magnetoelectric effects due to the Nambu-Goldstone modes have been identified in the dielectric spectrum of the Q1D spin-$\frac12$ multiferroic cuprate LiCu$_2$O$_2$. 
Further experiments on Q1D cuprates with weaker 3D couplings as in LiCuVO$_4$ \cite{naito:07,enderle:10} and Rb$_2$Cu$_2$Mo$_3$O$_{12}$ \cite{hase:04} might also uncover gapless dielectric spectra due to phasons and chirality solitons \cite{furukawa:08}.

The authors thank T. R\~{o}\~{o}m, H. H\"{u}vonen, S.-W. Cheong, T. Arima, Y. Tokura, Y. Yasui, A. Kobayashi, K. Okunishi, and A. Furusaki for discussions. 
The work was partly supported by 
Grants-in-Aid for Scientific Research (No.~19052006, No.~20029006, and No.~20046016) from MEXT of Japan and No. 21740275 from JSPS.


\newcommand{\etal}{{\it et al.}}

\newcommand{\PRL}[3]{Phys.\ Rev.\ Lett.\ {\bf #1}, #2 (#3)}
\newcommand{\PRLp}[3]{Phys.\ Rev.\ Lett.\ {\bf #1}, #2 (#3)}
\newcommand{\PRA}[3]{Phys.\ Rev.\ A {\bf #1}, #2 (#3)}
\newcommand{\PRAp}[3]{Phys.\ Rev.\ A {\bf #1}, #2 (#3)}
\newcommand{\PRB}[3]{Phys.\ Rev.\ B {\bf #1}, #2 (#3)}
\newcommand{\PRBp}[3]{Phys.\ Rev.\ B {\bf #1}, #2 (#3)}
\newcommand{\PRBR}[3]{Phys.\ Rev.\ B {\bf #1}, #2 (R) (#3)}
\newcommand{\PRBRp}[3]{Phys.\ Rev.\ B {\bf #1}, R#2 (#3)}
\newcommand{\arXiv}[1]{arXiv:#1}
\newcommand{\condmat}[1]{cond-mat/#1}
\newcommand{\JPSJ}[3]{J. Phys.\ Soc.\ Jpn.\ {\bf #1}, #2 (#3)}
\newcommand{\PTPS}[3]{Prog.\ Theor.\ Phys.\ Suppl.\ {\bf #1}, #2 (#3)}

\end{document}


\title{
Supplementary Material for \\``Chiral order and electromagnetic dynamics in 1D multiferroic cuprates"
}
\author{Shunsuke Furukawa}
\altaffiliation{Present address: Department of Physics, University of Toronto, Toronto, Ontario M5S 1A7, Canada}
\affiliation{Condensed Matter Theory Laboratory, RIKEN, Wako, Saitama 351-0198, Japan}
\author{Masahiro Sato}
\affiliation{Condensed Matter Theory Laboratory, RIKEN, Wako, Saitama 351-0198, Japan}
\author{Shigeki Onoda}
\affiliation{Condensed Matter Theory Laboratory, RIKEN, Wako, Saitama 351-0198, Japan}
\date{\today}
\maketitle

\section{iTEBD calculation}

\begin{figure*}[t]
\begin{center}
\includegraphics[width=0.7\textwidth]{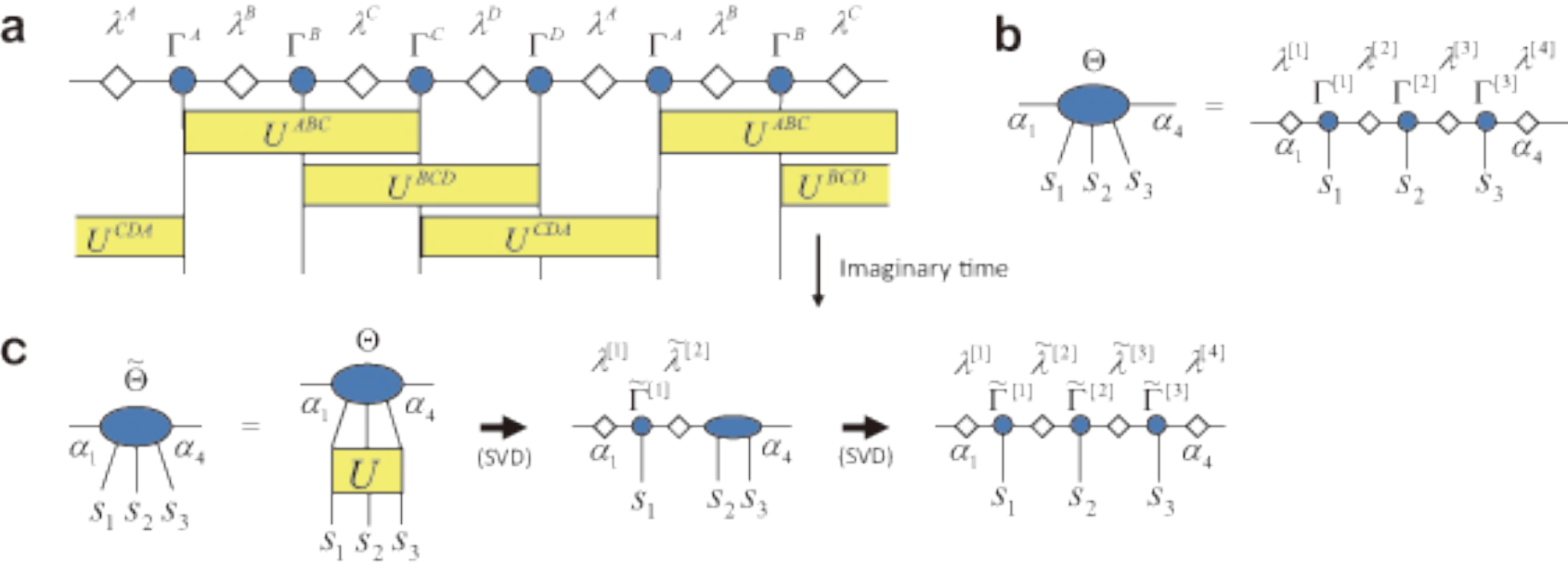}
\end{center}
\caption{(color) 
Processes in the present iTEBD algorithm.
(a) The period-4 matrix product state and its imaginary-time evolution. 
(b) Tensor $\Theta$ encapsulating the information of three sites $(1,2,3)$. 
(c) Two-step singular-value decompositions (SVD) of $\tilde{\Theta}$ carried out after operating a single imaginary-time evolution $U$.}
\label{fig:itebd}
\end{figure*}

In this section, we will give detailed explanations on the way how we apply the infinite time-evolving block decimation (iTEBD) algorithm, which has recently been developed by Vidal \cite{vidal:07}, for our precise determination of the phase diagram of the 1D spin-$\frac{1}{2}$ $J_1$-$J_2$ XXZ model. 

The method is based on a matrix product representation of a state, as is the density matrix renormalization group (DMRG) algorithm \cite{white:92}. A key feature of the iTEBD algorithm is that it directly treats an infinite system by exploiting the translational invariance, and is free from finite-size or boundary effects. 

A matrix product state (MPS) is an efficient and well-controlled variational ansatz for a 1D quantum ground state. We consider an infinite spin-$\frac12$ chain and define a local basis $|s_j\rangle~(s_j=\uparrow,\downarrow)$ for each site $j\in\mathbb{Z}$. To obtain a matrix product representation of a given state $|\Psi\rangle$, we first consider its Schmidt decomposition into the left and right half-infinite chains around a certain bond, for instance, $(0,1)$: 
\begin{equation}
 |\Psi\rangle = \sum_{\alpha=1}^\chi \lambda^{[1]}_\alpha |\Phi_\alpha^{[\lhd 0]}\rangle |\Phi_\alpha^{[1 \rhd]}\rangle.
\label{eq:Schmidt_decomp}
\end{equation}
Here, $[\lhd 0]$ and $[1\rhd]$ denote two half-infinite chains ranging over the sites with $j\le 0$ and $1\le j$ respectively,  and $\lambda^{[1]}_\alpha $ are the Schmidt coefficients controlling the entanglement between the two half chains. $\{ |\Phi_\alpha^{[\lhd 0]}\rangle \}$ and $\{ |\Phi_\alpha^{[1\rhd]}\rangle \}$ are the orthonormal sets of states on the left and right half chains, respectively. In general, an infinite sequence of Schmidt coefficients is required to exactly express a given state in an infinite system. Here, we introduce its approximate description by truncating the sequence: a finite number $\chi$ of the largest Schmidt coefficients are taken into account. 
A similar decomposition around the adjacent bond $(1,2)$ gives: 
\begin{equation}
 |\Psi\rangle = \sum_{\beta=1}^\chi \lambda^{[2]}_\beta |\Phi_\beta^{[\lhd 1]}\rangle |\Phi_\beta^{[2\rhd]}\rangle.
\label{eq: Schmidt_decomp2}
\end{equation}
Now one can consider the expansion of the state in $[1\rhd]$ using a local state at the site $1$ and the state in $[2\rhd]$:
\begin{equation}
 |\Phi_\alpha^{[1 \rhd]}\rangle 
 = \sum_{s_1,\beta} \Gamma^{[1]}_{\alpha s_1 \beta} \lambda^{[2]}_\beta 
   |s_1{}^{[1]}\rangle |\Phi_\beta^{[2\rhd]}\rangle.
\label{eq:Schmidt_decomp3}
\end{equation}
Here we have introduced a three-index tensor $\Gamma^{[1]}$. Plugging Eq.~(\ref{eq:Schmidt_decomp3}) into Eq.~(\ref{eq:Schmidt_decomp}), we obtain 
\begin{equation}
 |\Psi\rangle = \sum_{\alpha,s_1,\beta} 
 \lambda^{[1]}_\alpha \Gamma^{[1]}_{\alpha s_1 \beta} \lambda^{[2]}_\beta 
 |\Phi_\alpha^{[\lhd 0]}\rangle  |s_1{}^{[1]}\rangle |\Phi_\beta^{[2\rhd]}\rangle.
\end{equation}
Repeating this procedure over all the bonds, the state $|\Psi\rangle$ is eventually represented as an infinite product of vectors $\{\lambda^{[j]}_\alpha\}_{\alpha=1,\cdots,\chi}$ and three-index tensors $\{\Gamma^{[j]}_{\alpha,s_j,\beta}\}_{\alpha,\beta=1,\cdots,\chi;s_j=\uparrow,\downarrow}$ defined on the bond $(j-1,j)$ and the site $j$, respectively. The Schmidt rank $\chi$ controls the precision of the approximation, and plays a similar role to the number of states to be kept in the DMRG algorithm~\cite{white:92}. With larger $\chi$, one can describe longer-distance correlations more correctly. A simplification is achieved in the presence of the translational symmetry; tensors $\lambda^{[j]}$ and $\Gamma^{[j]}$ have a certain periodicity in $j$. For the analysis of the model $H_{1D}$ that includes nearest-neighbor and second-neighbor interactions, we employ a period-4 structure as in Fig.~S1(a). 

The calculation goes as follows. 
We first prepare an arbitrary initial state $|\Psi_0\rangle$ in the form of the MPS with the 4-lattice periodicity as shown in the top line of Fig.~S1(a). 
Then, we perform the imaginary-time evolution so that it converges to the ground state $|\Psi_{\rm GS}\rangle $: 
\begin{equation}
 |\Psi_{\rm GS}\rangle = \lim_{\tau\to \infty} \frac{e^{-H\tau} |\Psi_0\rangle}{ || e^{-H\tau} |\Psi_0\rangle || }
\end{equation}
In the present case, the imaginary-time evolution operator $e^{-H \Delta\tau}$ for a small time interval $\Delta\tau$ can be expanded through a Suzuki-Trotter decomposition as a sequence of three-site gates $\{U\}$ as shown in Fig.~S1(a). We now consider an operation of a single gate $U$ on three sites labeled by $(1,2,3)$. It is useful to rewrite the state as
\begin{equation}
 |\Psi\rangle = \!\!\! \sum_{\alpha_1,s_1,s_2,s_3,\alpha_4}  \!\!
 \Theta_{\alpha_1 s_1 s_2 s_3 \alpha_4}
 |\Phi^{[\lhd 0]}_{\alpha_0} \rangle |s_1^{[1]}\rangle |s_2^{[2]}\rangle |s_3^{[3]}\rangle 
 |\Phi^{[4 \rhd]}_{\alpha_4}\rangle
\end{equation}
with a short-hand notation on the coefficients [see Fig.~S1(b)],
\begin{equation}
 \Theta_{\alpha_1 s_1 s_2 s_3 \alpha_4} 
 = \!\! \sum_{\alpha_2,\alpha_3}
   \lambda_{\alpha_1}^{[1]} \Gamma_{\alpha_1 s_1 \alpha_2} ^{[1]}
   \lambda_{\alpha_2}^{[2]} \Gamma_{\alpha_2 s_2 \alpha_3} ^{[2]}
   \lambda_{\alpha_3}^{[3]} \Gamma_{\alpha_3 s_3 \alpha_4} ^{[3]}
   \lambda_{\alpha_4}^{[4]}.
\end{equation}
After the operation of $U$ on the sites (1,2,3), the coefficients are replaced by 
\begin{equation}
 \tilde{\Theta}_{\alpha_1 s_1 s_2 s_3 \alpha_4}
 = \sum_{\tilde{s}_1,\tilde{s}_2,\tilde{s}_3}
\langle s_1,s_2,s_3| U |\tilde{s}_1\tilde{s}_2 \tilde{s}_3 \rangle 
   \Theta_{\alpha_1 \tilde{s}_1\tilde{s}_2 \tilde{s}_3 \alpha_4}. 
\end{equation}
The original form of the MPS can be obtained by performing singular value decompositions twice for $\tilde{\Theta}$ [Fig.~S1(c)]. Here we truncate the tensors taking only the $\chi$ largest singular values (Schmidt coefficients) $\{\lambda_\alpha\}_{\alpha=1,\cdots,\chi}$ again. We repeat this imaginary-time evolution by taking the sites (1,2,3), (2,3,4), and (4,5,6), one after another [Fig.~S1(a)]. Since the same operator $U$ is applied for every four sites in parallel in an infinite chain at each imaginary time slice, the period-4 structure of the state is preserved.

\section{Determination of the phase diagram of the 1D model}

Here, we explain how we determine the phase diagram (Fig.~2(b) of the main body) for the 1D spin-$\frac12$ $J_1$-$J_2$ XXZ model, Eq.~(1) of the main body. 

A convergence to a vector-chiral long-range ordered state is allowed for by including complex elements in the initial MPS as a ``seed'' for the symmetry breaking \cite{okunishi:08}. The ``spontaneous'' chiral ordering can then be observed directly by calculating the local chirality 
\begin{equation} \label{eq:chiral_order}
 \kappa^z := \langle (\bm{S}_1\times\bm{S}_{2})^z \rangle. 
\end{equation}
It turned out that $\kappa^z$ is uniform and independent of the sites inside the chiral phase shown in Fig.~2(b) of the main body.
The chiral order parameter $\kappa^z$ calculated in this way for the model $H_{1D}$ with $J_1/J_2=-2$ is shown in Fig.~S2. 
The rapid growth of the order parameter allows us to determine the transition points with rather high accuracy. 
The $\chi$-dependence of the data is visible only in the close vicinity of the transition points. 
Therefore the calculation with $\chi=200$ is satisfactory for most of the parameter space, and this rank is employed in the calculation shown in Fig.~2(b) of the main text. 
On the antiferromagnetic side of $J_1$, the phase boundary determined in this way agrees well with the previous DMRG result \cite{hikihara:01}. 
In the XY case $\Delta=0$, the phase diagram is symmetric for $J_1<0$ and $J_1>0$ because the sign of $J_1$ can be reversed under the unitary transformation,
$W = \prod_j (2S_{2j}^z)$.

\begin{figure}[t]
\begin{center}
\includegraphics[width=0.45\textwidth]{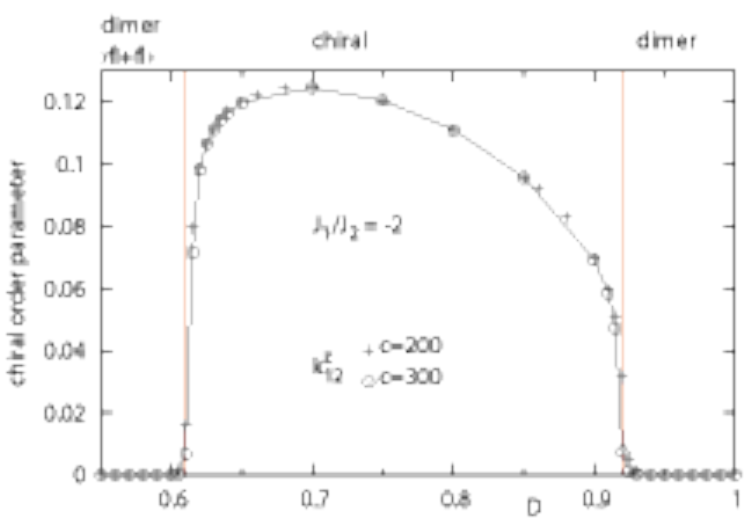}
\end{center}
\caption{
The chiral order parameter $\kappa^z$ for $J_1/J_2=-2$ of the 1D spin-$\frac12$ model, calculated by iTEBD with the ranks $\chi=200,~300$. 
The vertical lines indicate estimated transition points.
}
\label{fig:chiral_order}
\end{figure}

The dimer phases can be partly discriminated by calculating two types of dimer order parameters:
\begin{eqnarray}
 D_{123}^{xy} &=& (S_1^x S_2^x + S_1^y S_2^y) - (S_2^x S_3^x + S_2^y S_3^y),\\
 D_{123}^z    &=& S_1^z S_2^z - S_2^z S_3^z.
\end{eqnarray}
An alternation of the sign of $D_{123}^{xy}$ and/or $D_{123}^z$ along the spin chain indicates some sort of dimer ordering. 
In the singlet dimer phase for $J_1>0$, both $D_{123}^{xy}$ and $D_{123}^{z}$ are finite and have the same sign.
In particular, for $J_1/J_2=2$ and $\Delta>-1/2$, the ground state is exactly given by the product of nearest-neighbor singlets  
$(|\uparrow\downarrow\rangle-|\downarrow\uparrow\rangle)/\sqrt{2}$ \cite{majumdar:69}, which satisfies the relation $D_{123}^{xy}=2D_{123}^z= -1/2$. 
Applying the transformation $W$ in the XY case, one finds that the exact ground state for $J_1/J_2=-2$ and $\Delta=0$ is given by the dimer state whose unit is replaced by the triplet type, i.e., $(|\uparrow\downarrow\rangle+|\downarrow\uparrow\rangle)/\sqrt{2}$. 
This state satisfies the relation $-D_{123}^{xy}=2D_{123}^z= -1/2$. 
Our iTEBD calculation shows that $D_{123}^{xy}$ and $D_{123}^z$ always show opposite signs inside the dimer phase appearing in $\Delta\lesssim 0.7$ on the ferromagnetic side of $J_1$, indicating that such a ``triplet'' character of the wavefunction survives in this phase. 
On the other hand, in the other dimer phase close to the isotropic case $\Delta=1$ for $J_1<0$, both $D_{123}^{xy}$ and $D_{123}^z$ show finite values with the same sign, as in the singlet dimer phase appearing for $J_1>0$, though their amplitudes are significantly reduced, as suggested by field-theoretical analyses \cite{nersesyan:98, itoi:01}.

The Tomonaga-Luttinger liquid (TLL) phases appearing for $|J_1|/J_2\gtrsim4$ have the instability toward gapped antiferromagnetic (N\'eel) and dimer phases \cite{haldane:80, haldane:82}. 
The phase transitions to these two ordered phases can be analyzed efficiently by the level spectroscopy method \cite{nomura:94}, which combines the effective sine-Gordon theory with exact-diagonalization calculations. 
The phase boundary between the TLL phase and the dimer phase has been thus determined previously both for $J_1>0$ \cite{nomura:94} and for $J_1<0$ \cite{somma:01,fsf}. Besides, a N\'eel ordered phase with an antiferromagnetic long-range-order in the $z$ direction has been found between the TLL and chiral phases in $0.7\lesssim \Delta \lesssim 0.9$ \cite{fsf}.

\section{Crystal structure of LiCu$_2$O$_2$ and the spin model Hamiltonian}

Here, we explicitly give a mathematical expression of the model Hamiltonian for LiCu$_2$O$_2$.

LiCu$_2$O$_2$ has the space group $P_{nma}$ with the lattice constants $a=5.726$~\AA, $b=2.8587$~\AA, and $c=12.4137$~\AA, an inversion center, and a mirror plane perpendicular to the $b$ axis. 
For spin-$\frac12$, magnetic anisotropy is usually dominated by the Dzyaloshinskii-Moriya (DM) interaction. The symmetric anisotropic exchange coupling is of higher-order in the relativistic spin-orbit coupling.
Taking account of the symmetry properties of the crystal, we can construct the Hamiltonian including the DM interactions;
\begin{equation}
{\cal H} = \sum_{m=\mathrm{I}}^\mathrm{IV}{\cal H}^{2D}_m+{\cal H}_\perp+{\cal H}_{ac},
\label{eq:S:H}
\end{equation}
where
\begin{widetext}
\begin{eqnarray}
{\cal H}^{2D}_m &=& \sum_{i,j,\ell}\left[
J_1\bm{S}_{m,i,j,\ell}\cdot\bm{S}_{m,i,j+1,\ell}+
J_2\bm{S}_{m,i,j,\ell}\cdot\bm{S}_{m,i,j+2,\ell}
+\bm{D}_m\cdot\bm{S}_{m,i,j,\ell}\times\bm{S}_{m,i,j+1,\ell}+
J_a\bm{S}_{m,i,j,\ell}\cdot\bm{S}_{m,i+1,j,\ell}\right],
\ \ \ \ \ 
\label{eq:S:H2D}\\
{\cal H}_\perp &=& J_\bot\sum_{i,j,\ell}\left[
\left(\bm{S}_{\mathrm{II},i,j,\ell}+\bm{S}_{\mathrm{II},i,j+1,\ell}\right)\cdot\bm{S}_{\mathrm{III},i,j,\ell}+
\bm{S}_{\mathrm{IV},i+1,j,\ell+1}\cdot\left(\bm{S}_{\mathrm{I},i,j,\ell}+\bm{S}_{\mathrm{I},i,j+1,\ell}\right)
\right],
\label{eq:S:Hbc}\\
{\cal H}_{ac} &=& J'\sum_{i,j,\ell}\left[
\bm{S}_{\mathrm{I},i,j,\ell}\cdot\left(\bm{S}_{\mathrm{II},i-1,j,\ell}+\bm{S}_{\mathrm{II},i,j,\ell}\right)+
\left(\bm{S}_{\mathrm{III},i-1,j,\ell}+\bm{S}_{\mathrm{III},i,j,\ell}\right)\cdot\bm{S}_{\mathrm{IV},i,j,\ell}
\right]
\nonumber\\
&&+D'_b\sum_{i,j,\ell}\left[
\bm{S}_{\mathrm{I},i,j,\ell}\times\left(\bm{S}_{\mathrm{II},i-1,\ell}-\bm{S}_{\mathrm{II},i,\ell}\right)
+\left(\bm{S}_{\mathrm{III},i-1,j,\ell}-\bm{S}_{\mathrm{III},i,j,\ell}\right)\times\bm{S}_{\mathrm{IV},i,j,\ell}
\right]^b
\nonumber\\
&&+D''_a\sum_{\delta_a,\delta_b=\pm1}\delta_a\delta_b\left[\bm{S}_{\mathrm{I},i,j,\ell}\times\bm{S}_{\mathrm{II},i+\frac{1+\delta_a}{2},j+\delta_b,\ell}
+\bm{S}_{\mathrm{IV},i,j,\ell}\times \bm{S}_{\mathrm{III},i+\frac{1+\delta_a}{2},j+\delta_b,\ell} \right]^a
\nonumber\\
&&+D''_b\sum_{\delta_a,\delta_b=\pm1} \left[\bm{S}_{\mathrm{I},i,j,\ell}\times\bm{S}_{\mathrm{II},i+\frac{1+\delta_a}{2}, j+\delta_b,\ell}
+\bm{S}_{\mathrm{IV},i,j,\ell}\times \bm{S}_{\mathrm{III},i+\frac{1+\delta_a}{2},j+\delta_b,\ell}\right]^b
\nonumber\\
&&+D''_c\sum_{\delta_a,\delta_b=\pm1} \delta_b\left[\bm{S}_{\mathrm{I},i,j,\ell}\times\bm{S}_{\mathrm{II},i+\frac{1+\delta_a}{2},j+\delta_b,\ell}
+\bm{S}_{\mathrm{IV},i,j,\ell}\times \bm{S}_{\mathrm{III},i+\frac{1+\delta_a}{2},j+\delta_b,\ell}\right]^c.
\label{eq:S:Hac}
\end{eqnarray}
\end{widetext}
Here, $\bm{S}_{m,i,j,\ell}$ represents the electronic spin located at the position $\bm{r}_{m,i,j,\ell}$ of the Cu$^{2+}$ ion, with
\begin{align}
\bm{r}_{\mathrm{I},i,j,\ell}&=((i+0.12)a,(j+0.25)b,(\ell+0.905)c),
\\
\bm{r}_{\mathrm{II},i,j,\ell}&=(i+0.62)a,(j+0.25)b,(\ell+0.595)c),
\\
\bm{r}_{\mathrm{III},i,j,\ell}&=(i+0.38)a,(j+0.75)b,(\ell+0.405)c),
\\
\bm{r}_{\mathrm{IV},i,j,\ell}&=(i-0.12)a,(j+0.75)b,(\ell+0.095)c).
\end{align}
The Heisenberg exchange couplings $J_1$, $J_2$, $J_a$, $J_\perp$, and $J'$, the intra-chain DM vectors $\bm{D}_{I,\cdots,IV}$ specified by $D$ and $\theta$, and the inter-chain DM couplings $D''_a$, $D''_b$, and $D''_c$ have been explained in the main text. 
A finite but rather weak nearest-neighbor inter-chain DM vector $(0,D'_b,0)$ may also exist because of a slight shift of the atoms from the inversion-symmetric positions about the bond center. Most likely, however, the second-neighbor inter-chain DM couplings $D''_{a,b,c}$ [Fig.~1(e) of the main body] prevail over $D'_b$, since the inversion symmetry is broken more significantly in this geometry and the relevant exchange processes include common paths. In particular, $D''_a$ is the only DM coupling that favors the uniform $bc$ spin-spiral component observed in experiments~\cite{park:07,seki:08,kobayashi:09,huvonen:09}, and should play a central role in fixing the spiral planes. To reduce the number of parameters, we neglect $J'$, $D'_b$, $D''_b$, and $D''_c$ for LiCu$_2$O$_2$. Actually, the other components favors alternating spiral structures as proposed for NaCu$_2$O$_2$~\cite{capogna:05}.
Note that the existence of the inversion center in the bonds connecting zigzag-ladder pairs, namely, between the layers II and III and between the layers I and IV, prohibits the DM interaction on these bonds. We ignore the DM interaction on the bonds directed to the $a$ axis, since it is small compared with $J_a$ and does not stabilize the experimentally observed staggered spin modulation along the $a$ axis.

\section{Classical analyses of magnetic order}

Here, we explain details of the classical analysis of the magnetic structure for LiCu$_2$O$_2$, whose result is shown in Fig.~1(f) of the main body. We ignore an ellipticity caused by a density wave of magnons, which is beyond the scope of a linear spin-wave theory.

We assume the magnetic structure of 
$\langle\bm{S}_{m,i,j,\ell}\rangle=S\bm{n}_{m,i,j,\ell}$,
where $\bm{n}_{m,i,j,\ell}=(n^x_{m,i,j,\ell},n^y_{m,i,j,\ell},n^z_{m,i,j,\ell})$ is a unit vector whose components are given by
\begin{subequations}
\begin{align}
&n^x_{m,i,j,\ell}=\cos\psi_m\sin\varphi_m\cos\theta_{m,i,j,\ell}-\sin\psi_m\sin\theta_{m,i,j,\ell},
\label{eq:S:n^x}\\
&n^y_{m,i,j,\ell}=\sin\psi_m\sin\varphi_m\cos\theta_{m,i,j,\ell}+\cos\psi_m\sin\theta_{m,i,j,\ell},
\label{eq:S:n^y}\\
&n^z_{m,i,j,\ell}=\cos\varphi_m \cos\theta_{m,i,j,\ell}.
\label{eq:S:n^z}
\end{align}
\label{eq:S:n}
\end{subequations}
Here, $\varphi_m$ and $\psi_m$ represent the successive rotation angles of the cycloid plane in the $m$th layer about the $b$ axis from the $bc$ plane and then about the $c$ axis, respectively. We have introduced the phase $\theta_{m,i,j,\ell}$ of the spin spiral at the site $\bm{r}_{m,i,j,\ell}$ in the $m$th chain as
\begin{align}
\theta_{\mathrm{I},i,j,\ell}&=\bm{Q}\cdot(ia,jb,\ell c)+\phi_I,
\\
\theta_{\mathrm{II},i,j,\ell}&=\bm{Q}\cdot((i+1/2)a,jb,\ell c)+\phi_{II},
\\
\theta_{\mathrm{III},i,j,\ell}&=\bm{Q}\cdot((i+1/2)a,(j+1/2)b,\ell c)+\phi_{III},
\\
\theta_{\mathrm{IV},i,j,\ell}&=\bm{Q}\cdot(ia, (j+1/2)b,\ell c)+\phi_{IV}.
\end{align}
We substitute these expressions into the Hamiltonian Eq.~(\ref{eq:S:H}) with $D_b'=D_b''=D_c''=0$, and minimize it with respect to $\bm{Q}$, $\varphi_m$, $\psi_m$, and $\phi_m$ ($m=I,\cdots,IV$). 
Note that large values of $D_a''$ and $J_a>0$ compared with $|J'|$ are required for reproducing the experimentally observed ordering wavevector $\bm{Q}=(Q_a,Q_b,0)$ with $Q_a=\pi/a$~\cite{seki:08,kobayashi:09, masuda:04,masuda:05}. In fact, once $Q_a=\pi/a$ is obtained, small $J'$ does not alter the magnetic structure at all. Hence, we take $J'=0$ for simplicity.

Then, the minimization procedure gives generic conditions for the rotation angles, $\psi_m=0$, $\varphi_I=-\varphi_{III}=\varphi_1$, and $\varphi_{II}=-\varphi_{IV}=\varphi_2$, and those for the initial phases $\phi_{II}-\phi_I=-\pi/2$, $\phi_{III}-\phi_I=\pi/2$, and $\phi_{IV}-\phi_I=\pi$, as described in the text and Fig.~1(f) of the main body.

\section{Determination of the exchange and DM coupling parameters}

Here, we explain in detail how we have determined the set of parameter values $(J_1, J_2, J_a, J_\perp, D, D''_a) =(-11.3, 5.9, 1.1, 0.08, 0.37, 0.26)$ meV, and $\theta =0.43\times2\pi$ from a fitting with experimental results.

\begin{figure}[b]
\begin{center}
\includegraphics[width=0.45\textwidth]{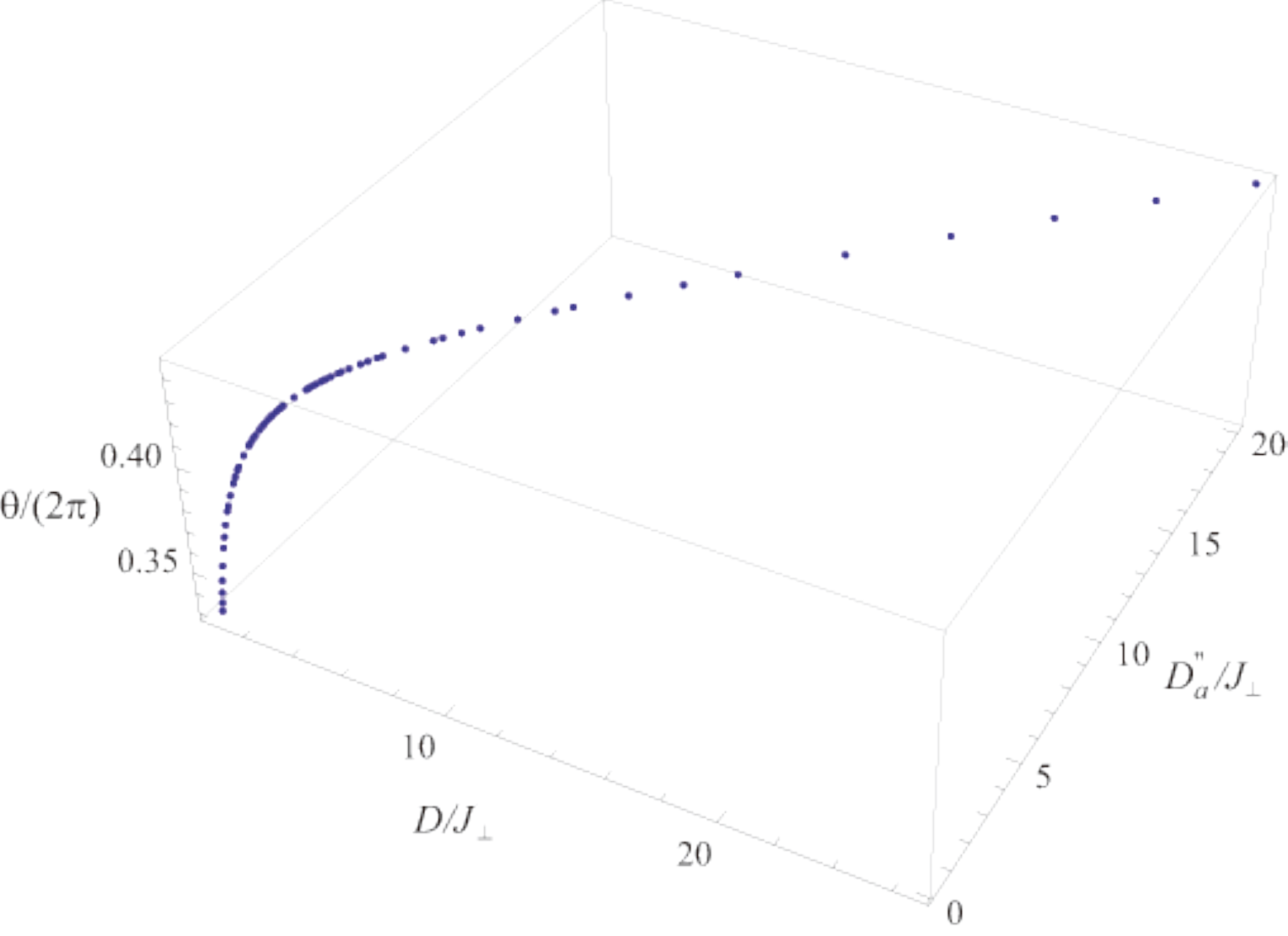}
\end{center}
\caption{
The solutions of $(D/J_\perp,D''_a/J_\perp,\theta)$ for which the classical ground state satisfies the condition $(\varphi_1,\varphi_2)/(2\pi)=(-0.08,-0.02)$ required for fitting the NMR spectra [Fig.~\ref{fig:S:NMR}].
}
\label{fig:S:DetermineFromNMR}
\end{figure}

(i) We tune $J_2/J_1$ to reproduce the neutron-scattering result $Q_b=0.173\times2\pi/b$~\cite{ seki:08,kobayashi:09,masuda:04,masuda:05}. 

(ii) As shown for the present case of $Q_b=0.173\times2\pi/b$ in Fig.~\ref{fig:S:DetermineFromNMR}, there exists a single curve in the space of $(D/J_\perp,\theta,D_a''/J_\perp)$ that reproduces $(\varphi_1,\varphi_2)/(2\pi)=(-0.08,-0.02)$, which is determined so as to fit the $^7$Li NMR spectra~\cite{kobayashi:09,gippius:04,svistov:09} (see the next section). Thus, $D/J_\perp$ and $\theta$ are fixed to satisfy this condition. 

(iii) The rest of the parameters, i.e., $J_1$, $J_a$, $J_\perp$, and $D_a''$, are determined to explain the experimental results on the anisotropy in the antiferromagnetic resonance spectra \cite{huvonen:09}, $\mathrm{Im}~\delta\chi(\omega)=\mathrm{Im}~[\chi^{cc}(\omega)-(\chi^{aa}(\omega)+\chi^{bb}(\omega))/2]$. The energy levels of the modes observed with the THz spectroscopy~\cite{huvonen:09} are given by $E_A=0.83$, $E_B=1.04$, $(E_{C1},E_{C2})=(1.43,1.51)$, $E_D=2.39$, and $(E_{E1},E_{E2})=(2.98,3.06)$ in meV. We tune $J_1$, $J_a$, $J_\perp$, and $D_a''$ to reproduce the energy levels and the intensities of three main peaks observed for $\mathrm{Im}~\delta\chi(\omega)$ at $\omega=E_A$, $E_{C1}$, and $E_{E1}$ \cite{huvonen:09} [Fig.~3(b) of the main body].

\section{$^7$Li NMR spectra for LiCu$_2$O$_2$}

Two parameters $\varphi_1$ ($=\varphi_I=-\varphi_{III}$) and $\varphi_2$ ($=\varphi_{II}=-\varphi_{IV}$) of the magnetic ordering pattern are fixed to reproduce the $^7$Li NMR spectra for LiCu$_2$O$_2$ \cite{kobayashi:09, gippius:04,svistov:09}, as follows. 

In a magnetically long-range ordered phase, the $^7$Li NMR spectra could be dominated by the dipole hyperfine field due to the ordered spin moments;
\begin{widetext}
\begin{equation}
\bm{h}_{\mathrm{dip}}(\bm{r}_{\mathrm{Li}})=\sum_{m,i,j,\ell}(-g\mu_B)\left[3\frac{\left(\bm{S}_{m,i,j,\ell}\cdot\left(\bm{r}_{m,i,j,\ell}-\bm{r}_{\mathrm{Li}}\right)\right)\left(\bm{r}_{m,i,j,\ell}-\bm{r}_{\mathrm{Li}}\right)}{ |\bm{r}_{m,i,j,\ell}-\bm{r}_{\mathrm{Li}}|^5}-\frac{\bm{S}_{m,i,j,\ell}}{|\bm{r}_{m,i,j,\ell}-\bm{r}_{\mathrm{Li}}|^3}\right],
\label{eq:S:h_dip}
\end{equation}
\end{widetext}
with the $g$ factor of electrons, the Bohr magneton $\mu_B$, the electronic spin $\bm{S}_{m,i,j,\ell}$ at the Cu$^{2+}$ position $\bm{r}_{m,i,j,\ell}$, and the position $\bm{r}_{\mathrm{Li}}$ of a Li nuclei. The NMR spectrum at the static applied magnetic field $\bm{H}$ is calculated through 
\begin{equation}
I(\bm{H},\nu)=\mathop{\mathrm{Ave}}_{{\bm{r}_{\mathrm{Li}}}}~\delta\left(\frac{\nu}{2\pi}-\gamma_{{}^7\mathrm{Li}}|\bm{H}+\bm{h}_{\mathrm{dip}}(\bm{r}_{\mathrm{Li}})|\right),
\label{eq:S:NMR}
\end{equation}
with the gyromagnetic ratio for the $^7$Li nuclear magnetic moment, $\gamma_{{}^7\mathrm{Li}}$. 

\begin{figure}[b]
\begin{center}
\includegraphics[width=0.4\textwidth]{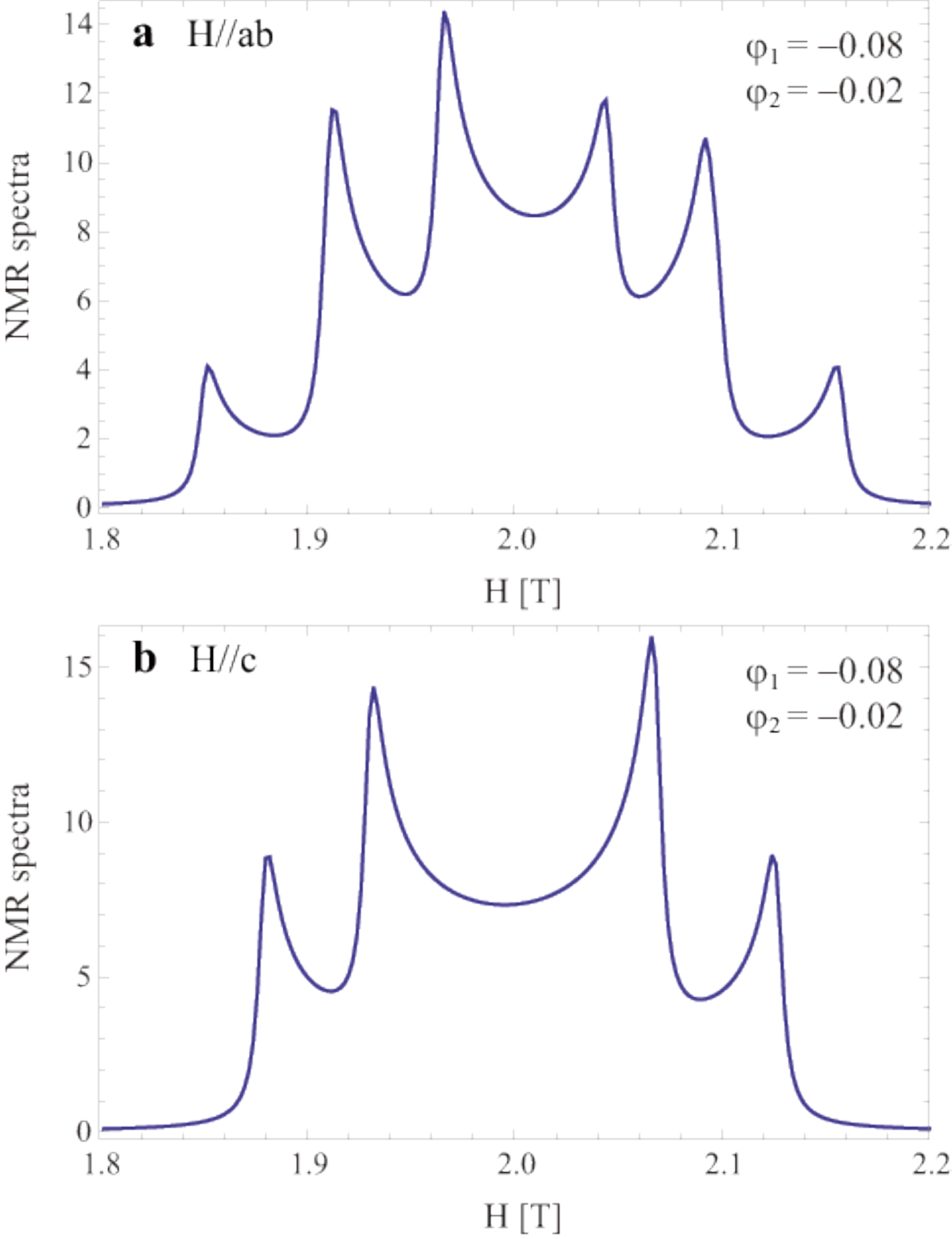}
\end{center}
\caption{
The $^7$Li NMR spectra for the $ab$-twinned LiCu$_2$O$_2$ as functions of the dc magnetic field $H$ at the ac magnetic field with the frequency $\nu=33.4$ MHz for $(\varphi_1,\varphi_2)= (-0.08,-0.02)$.
(a) The spectrum averaged over the $\bm{H}\parallel a$ and $\bm{H}\parallel b$ configurations.
(b) The spectrum for $\bm{H}\parallel c$.}
\label{fig:S:NMR}
\end{figure}

Fittings to the $^7$Li NMR spectra at the static in-plane ($ab$-plane) and out-of-plane magnetic fields are performed by changing $\varphi_1/(2\pi)$ and $\varphi_2/(2\pi)$ by $1\%$. Then, several different sets of the rotation angles, $(\varphi_1,\varphi_2)/(2\pi)\sim(-0.03,0.04)$, $(-0.08,0.10)$, and $(-0.08,-0.02)$, reasonably reproduce the experimental results by Gippius {\it et al.}~\cite{gippius:04}, as shown for the case of $(\varphi_1,\varphi_2)/(2\pi)= (-0.08,-0.02)$ in Fig.~\ref{fig:S:NMR}. We have not introduced an additional adjustable parameter for the transferred hyperfine coupling, as in the previous analyses~\cite{kobayashi:09,gippius:04}. The magnetic patterns of the first and the second sets with $\varphi_1\approx-\varphi_2$ are not stabilized in our case of $D_a''\ne0$. Thus, we take the last set $(\varphi_1,\varphi_2)/(2\pi)= (-0.08,-0.02)$.

\section{Linear spin-wave theory for LiCu$_2$O$_2$}

Following the conventional linear spin-wave approximation for the helical magnetism \cite{nagamiya}, we introduce the Holstein-Primakoff bosons for the spin-$S$ operators $\bm{T}_{\bm{r}}=(T^x_{\bm{r}},T^y_{\bm{r}},T^z_{\bm{r}})$ at the site $\bm{r}$ as
\begin{subequations}
\begin{eqnarray}
T^x_{\bm{r}}+i T^y_{\bm{r}} &=& \sqrt{2S}\left[1-\frac{1}{2S}b^\dagger_{\bm{r}}b_{\bm{r}}\right]^{-\frac{1}{2}}b_{\bm{r}},
\label{eq:S:T+}\\
T^x_{\bm{r}}-i T^y_{\bm{r}} &=&\sqrt{2S} b^\dagger_{\bm{r}}\left[1-\frac{1}{2S}b^\dagger_{\bm{r}}b_{\bm{r}}\right]^{-\frac{1}{2}},
\label{eq:S:T-}\\
T^z_{\bm{r}}&=&S-b^\dagger_{\bm{r}}b_{\bm{r}}.
\label{eq:S:Tz}
\end{eqnarray}
\label{eq:S:Ttob}
\end{subequations}
We rotate $\bm{T}_{\bm{r}}$ so that its $z$ component represents the predicted magnetic structure and its $x$ component is perpendicular to the spiral plane. Thus, we perform the unitary transformation $\bm{S}_{\bm{r}}=U_{\bm{r}}\bm{T}_{\bm{r}}$ to obtain
\begin{subequations}
\begin{align}
S^x_{m,i,j,\ell} =&~ T^x_{ m,i,j,\ell }\cos\varphi_m 
- T^y_{ m,i,j,\ell }\sin\varphi_m\sin\theta_{ m,i,j,\ell } \notag\\
&+ T^z_{ m,i,j,\ell }\sin\varphi_m\cos\theta_{ m,i,j,\ell },
\label{eq:S:Sx}\\
S^y_{m,i,j,\ell} =&~ T^y_{ m,i,j,\ell }\cos\theta_{ m,i,j,\ell } 
+ T^z_{ m,i,j,\ell } \sin\theta_{ m,i,j,\ell },
\label{eq:S:Sy}\\
S^z_{m,i,j,\ell} =&~ -T^x_{ m,i,j,\ell }\sin\varphi_m 
- T^y_{ m,i,j,\ell }\cos\varphi_m\sin\theta_{ m,i,j,\ell } \notag\\
&+ T^z_{ m,i,j,\ell }\cos\varphi_m\cos\theta_{ m,i,j,\ell }.
\label{eq:S:Sz}
\end{align}
\label{eq:S:StoT}
\end{subequations}
We substitute Eqs.~(\ref{eq:S:Ttob}) into Eqs.~(\ref{eq:S:StoT}) and then Eqs.~(\ref{eq:S:StoT}) to the Hamiltonian Eq.~(\ref{eq:S:H}), and take only the bilinear terms in the bosons.

\begin{table*}[tb]
\begin{center}
\begin{tabular}{|c|c|c|c|c|c|}
\hline
Mechanism & Magnon & Intra-chain & \multicolumn{3}{|c|}{Inter-chain}
\\ \cline{4-6}
& & ($\alpha_{\kappa; 1}$) &$a$-axis ($\alpha_{\kappa,\mathrm{ms}; a}$) & II-III, IV-I ($\alpha_{\kappa; \perp}$) & I-II, III-IV ($\alpha_{\kappa}'$)
\\ \hline
Chirality & $\bm{k}=0$ & NA & NA & $E^\omega\parallel a,b,c$ & $E^\omega \parallel c$
\\ \cline{2-6}
($\alpha_{\kappa; \mu}$)& $\bm{k}=\pm\bm{Q}$ & $E^\omega \parallel a,c$ & NA & $E^\omega \parallel a,b,c$ & $E^\omega \parallel b,c$
\\ \hline
Magnetostriction & $\bm{k}=0$ & NA & NA & NA & NA
\\ \cline{2-6}
($\alpha_{\mathrm{ms}; \mu}$)& $\bm{k}=\pm\bm{Q}$ & NA & NA & NA & $E^\omega \parallel a,c$
\\ \hline
\end{tabular}
\end{center}
\caption{\label{tab:selction}
The selection rules for various contributions from bonds to the electric-dipole ($\bm{E}^\omega$) excitation through the chirality and magnetostriction mechanisms. The rows for ``Intra-chain'', ``$a$-axis'', ``II-III, IV-I'', and ``I-II, III-IV'' represent the contributions from the bonds corresponding to $J_1$ (also $J_2$), $J_a$, $J_\perp$, and $J'$ (also $D''_a$), respectively. They can have independent magnetoelectric couplings $\alpha_{\kappa,\mathrm{ms};\mu}$. ``NA'' means no activity, i.e., $\alpha_{\kappa,\mathrm{ms};\mu}=0$.}
\label{table:S:dipole}
\end{table*}

In the Fourier-space representation of the spin-wave Hamiltonian, there exist bilinear mixing terms of bosons $b_{m,\bm{q}}$ and $b^\dagger_{m,\bm{q}}$ with the wavevectors $\bm{q}$ and $\bm{q}\pm\bm{Q}$. They generate higher harmonics in the magnetic structure and activate the otherwise silent modes for the optical electric-dipole transitions. However, the mixing terms are proportional to inter-layer couplings or DM couplings, and hence are small compared with the intra-layer terms. Therefore, the modification of the spin-wave spectrum becomes important only when the energy level of the 2D spin-wave dispersion almost vanishes, namely, around $\bm{q}=0$ and $\pm\bm{Q}$ [circles in Fig.~3(a) of the main body]. This allows us to treat these mixing terms along the concept of the degenerate perturbation theory of $\bm{q}$ and $\bm{q}\pm\bm{Q}$ modes for $\bm{q}\approx0$. Namely, we take the basis of 
\begin{equation}
\begin{split}
\Psi_{\bm{q}}={}^t (\{b_{m,\bm{q-Q}},b_{m,\bm{-q+Q}}^\dagger,b_{m,\bm{q}},b_{m,-\bm{q}}^\dagger,\\
b_{m,\bm{q+Q}},b_{m,\bm{-q-Q}}^\dagger\}_{m=I,\cdots,IV}),
\end{split}
\label{eq:Psi}
\end{equation}
and express the linearized Hamiltonian as $\Psi_{\bm{q}}^\dagger H^{\mathrm{LSW}}_{\bm{q}} \Psi_{\bm{q}}$ with $\bm{q}\approx0$.
Then, we perform the Bogoliubov transformation by solving the generalized eigenvalue problem, 
\begin{eqnarray}
GH^{\mathrm{LSW}}_{\bm{q}} B_{\bm{q}} &=& B_{\bm{q}} E,
\\
B^\dagger_{\bm{q}} G B_{\bm{q}} &=& G,
\end{eqnarray}
where $G$ and $E$ are the $24\times24$ diagonal matrices whose elements are given by $(1,-1,1,-1,\cdots)$ and pairs of levels, $(\omega_1,-\omega_1,\omega_2,-\omega_2,\cdots)$, respectively. Then, we can describe the spin-wave eigenmodes in terms of the original spins by using the inverse Bogoliubov transformation $B^{-1}_{\bm{q}}$ and the Fourier transform of the inverse of the unitary transformation $U_{\bm{r}}$.

\section{Magnetoelectric couplings and the optical selection rules for LiCu$_2$O$_2$}

Here, we summarize the selection rules for local magnetoelectric couplings in various bonds through both contributions from the vector spin chirality and the magnetostriction to the electric dipole moment~\cite{moriya:67,jonh}, 
\begin{eqnarray}
P^i_{\kappa}&=&\sum_{\bm{r},\bm{r}'}\alpha^i_{\kappa}(\bm{r},\bm{r}')\left[\frac{\bm{r}-\bm{r}'}{|\bm{r}-\bm{r}'|}\times(\bm{S}_{\bm{r}}\times\bm{S}_{\bm{r}'})\right]^i,
\label{eq:S:IDM}
\\
P^i_{\mathrm{ms}}&=&\sum_{\bm{r},\bm{r}'}\alpha^i_{\mathrm{ms}}(\bm{r},\bm{r}') (\bm{S}_{\bm{r}}\cdot\bm{S}_{\bm{r}'}),
\label{eq:S:MS}
\end{eqnarray}
respectively, with $i$ being the index for the spatial direction. The coefficients $\alpha^i_{\kappa}(\bm{r},\bm{r}')$ and $\alpha^i_{\mathrm{ms}}(\bm{r},\bm{r}')$ depend on the details of both electronic and ionic polarizability, and quantitative microscopic calculations of these coefficients are beyond our scope. However, the contributions from various bonds can be analyzed from the symmetry consideration. The results are summarized in Table~\ref{table:S:dipole}.  

As far as the one-magnon contributions are concerned, the magnetostriction mechanism is active only for the magnons at the $\bm{q}=\pm\bm{Q}$ but not at the Brillouin zone boundary in the case of LiCu$_2$O$_2$, because the primitive unit cell contains only one Cu$^{2+}$ ion in both the $a$ and $b$ directions. Then, this mechanism works only in the bonds connecting the layers I and II and those connecting III and IV, having a negligibly small exchange coupling $J'$ between the nearest-neighbor spins. Therefore, the magnetoelectric coupling through the exchange-striction on these bonds should also be negligibly small.

In contrast, the chirality contributions are generally effective except the bonds along the $a$ axis. In particular, contributions from the inter-layer bonds between the layers II and III and between I and IV are geometrically advantageous from the viewpoint of the quantum chemistry. In the first order in the relativistic spin-orbit coupling, it can flip the spin at the Cu$^{2+}$ site in one chain, and simultaneously change the orbital shape from $d_{x^2-y^2}$ to $d_{yz}$ or $d_{zx}$. These $t_{2g}$ orbitals form $dp\pi$ bonds with the O2p orbital located along the $c$ axis, thus allowing for a virtual electron transfer between them. This O2p orbital also has a $dp\sigma$ hybridization with the $d_{x^2-y^2}$ orbital at the Cu$^{2+}$ site in the other chain. In this way, the inverse Dzyaloshinskii-Moriya mechanism works efficiently. 
 Therefore we took into account only these chirality contributions and adjusted the coefficients in Eq.~\eqref{eq:S:IDM} in order to reasonably reproduce the experimental data on the anisotropy in the dielectric function obtained from the THz spectroscopy~\cite{huvonen:09}.


\newcommand{\etal}{{\it et al.}}
\newcommand{\PRL}[3]{Phys.\ Rev.\ Lett.\ {\bf #1}, \href{http://link.aps.org/abstract/PRL/v#1/e#2}{#2} (#3)}
\newcommand{\PRLp}[3]{Phys.\ Rev.\ Lett.\ {\bf #1}, \href{http://link.aps.org/abstract/PRL/v#1/p#2}{#2} (#3)}
\newcommand{\PRA}[3]{Phys.\ Rev.\ A {\bf #1}, \href{http://link.aps.org/abstract/PRA/v#1/e#2}{#2} (#3)}
\newcommand{\PRAp}[3]{Phys.\ Rev.\ A {\bf #1}, \href{http://link.aps.org/abstract/PRA/v#1/p#2}{#2} (#3)}
\newcommand{\PRB}[3]{Phys.\ Rev.\ B {\bf #1}, \href{http://link.aps.org/abstract/PRB/v#1/e#2}{#2} (#3)}
\newcommand{\PRBp}[3]{Phys.\ Rev.\ B {\bf #1}, \href{http://link.aps.org/abstract/PRB/v#1/p#2}{#2} (#3)}
\newcommand{\PRBR}[3]{Phys.\ Rev.\ B {\bf #1}, \href{http://link.aps.org/abstract/PRB/v#1/e#2}{#2} (R) (#3)}
\newcommand{\PRBRp}[3]{Phys.\ Rev.\ B {\bf #1}, \href{http://link.aps.org/abstract/PRB/v#1/e#2}{R#2} (#3)}
\newcommand{\arXiv}[1]{arXiv:\href{http://arxiv.org/abs/#1}{#1}}
\newcommand{\condmat}[1]{cond-mat/\href{http://arxiv.org/abs/cond-mat/#1}{#1}}
\newcommand{\JPSJ}[3]{J. Phys.\ Soc.\ Jpn.\ {\bf #1}, \href{http://jpsj.ipap.jp/link?JPSJ/#1/#2/}{#2} (#3)}
\newcommand{\PTPS}[3]{Prog.\ Theor.\ Phys.\ Suppl.\ {\bf #1}, \href{http://ptp.ipap.jp/link?PTPS/#1/#2/}{#2} (#3)}
\newcommand{\hreflink}[1]{\href{#1}{#1}}

\newcommand{\bibitemex}[1]{
\stepcounter{count}
\bibitem[S\thecount]{#1}
}
\newcounter{count}